\documentclass[prl, twocolumn, showpacs, reprint, nofootinbib, superscriptaddress, floatfix]{revtex4-2}

\usepackage{graphicx}
\usepackage{dcolumn}
\usepackage{bm}
\usepackage{amssymb,amsmath,amsfonts}
\usepackage{graphics,epsfig,subfigure}
\usepackage{color,bm}
\usepackage{xcolor}
\usepackage{verbatim}
\usepackage{lipsum}

\newcommand{\ave}[1]{\left\langle #1 \right\rangle}

\begin{document}

\preprint{APS/123-QED}

\title{Critical Scaling of Finite-Size Fluctuations around Marginal Stability in Long-Range Hamiltonian Systems}
\author{Yoshiyuki Y. Yamaguchi}
\email{yyama@amp.i.kyoto-u.ac.jp}
\affiliation{
  Graduate School of Informatics, Kyoto University, Kyoto 606-8501, Japan}
\author{Julien Barr{\'e}}
\email{julien.barre@univ-orleans.fr}
\affiliation{
  Institut Denis Poisson, Universit{\'e} d'Orl{\'e}ans, Universit{\'e} de Tours and CNRS, 45067 Orl{\'e}ans, France}

\date{\today}

\begin{abstract}

  Finite size fluctuations are a crucial ingredient in kinetic theory
  of long-range interacting collisionless systems. In this Letter,
  we introduce a phenomenological theory which predicts an anomalous
  scaling close to marginal stability for these fluctuations. It also
  pinpoints the critical window inside which the fluctuations are
  anomalous, and outside which they are Gaussian. Shrinking very
  slowly as $N^{-1/5}$, this critical window encompasses a wide region
  around marginal stability. We confirm our predictions through
  extended numerical simulations on two different simplified models. 

\end{abstract}

\maketitle

\paragraph{Introduction.}
Hamiltonian collisionless partial differential equations (PDE) appear
in many different domains across physical sciences:
Collisionless Boltzmann equation for self gravitating systems \cite{BinneyTremaine},
Vlasov-Poisson equation for plasmas \cite{Nicholson83},
and Euler equation for fluids \cite{Morrison82}.
In many cases, the collisionless PDE is actually a continuum approximation
of a microscopic dynamics involving $N$ particles interacting at long range:
self gravitating stars, electrons and ions in electrostatic interactions,
and point vortices for instance. 
This large $N$ limit, i.e. from $N$-body Hamiltonian dynamics to a Vlasov-like PDE,
is rationalized by mathematical theorems, at least over a finite time horizon
\cite{Dobrushin,Neunzert,BraunHepp,Hauray07,Hauray15,Lazarovici17}.
Finite $N$ fluctuations  around this deterministic evolution are physically crucial,
since they drive the secular evolution of these systems,
over time scales of order $N$;
see \cite{Balescu,Lenard}, and \cite{Feliachi22}
for large deviations. 

These finite size fluctuations are usually described by a central limit theorem (CLT),
entailing Gaussian fluctuations of order $N^{-1/2}$, see \cite{BraunHepp,Lancellotti09}.
Close to a stable stationary state, these fluctuations then behave at large time
as a stationary Gaussian process, which has long been known in the physics literature
(see for instance \cite{Lifschitz_book}), and which is studied rigorously in \cite{Lancellotti16}.

However, the standard CLT-like description of fluctuations
relies on a well-behaved linearized dynamics,
and hence is valid only far from any instability threshold.
What are the near critical fluctuations of these interacting particles systems?
How do they scale with $N$ and the distance to instability?
We address these widely open questions in this article and uncover new scaling laws.
 
To be more precise,
we concentrate on the vicinity of a particular type of bifurcation for Vlasov-like equations,
characterized by "trapping scaling"
(the saturated amplitude of the growing unstable mode is $O(\lambda^{2})$,
where $\lambda$ is the real growth rate),
and whose normal form near marginal stability is a PDE
called the "Single Wave Model" (SWM) \cite{ONeil71}.
The nonlinear dynamics close to this bifurcation is dominated
by the interaction between the marginally stable mode and particles with similar phase velocity,
whose mathematical signature is a continuous purely imaginary spectrum:
a critical layer at this resonant velocity forms,
and a structure sometimes called a "cat's eye" develops.
The width of the cat's eye is typically $O(|A|^{1/2})$
where $A$ is the complex amplitude of the weakly unstable mode.
At the same time, the dynamics is strongly constrained
by the infinite number of conserved quantities
stemming from the degenerate Hamiltonian structure of the PDE
(these are the Casimirs of the theory).
These features create a universal bifurcation structure,
with specific exponents \cite{Ivanov01,Ivanov05,YoshiShun14,YoshiShun15}. 
This bifurcation, ubiquitous in Vlasov-type PDEs,
describes the saturation of instabilities in plasmas \cite{Dewar73},
vortex formation in critical layers in fluid dynamics \cite{Churilov},
and has been used to explain the saturation amplitude of spiral arms
in galactic dynamics \cite{Morozov80,Hamilton_spiral}:
this diversity of applications attests the universality of this bifurcation,
emphasized explicitly in \cite{DelCastilloNegrete,BMT}.

It is natural to study the amplitude $A(t)$ of the near marginal mode,
which plays the role of an "order parameter" for the SWM bifurcation.
We shall ask the following questions: 
(i) {\it What is the variance $V$ of $|A(t)|$}
at the critical point? It scales as $V=O(N^{-\rho})$, and
CLT arguments would lead to $\rho=1$ in general.
Close to the critical point, mean field critical fluctuations,
which could be thought of as a reasonable guess for long range interacting systems,
would lead to $\rho=1/2$;
seminal simulations \cite{Yoshi2016} on the other hand suggests an anomalous exponent
close to $\rho=4/5$.
(ii) {\it What is the probability distribution function (PDF) of the order parameter?}
It is expected to be approximately Gaussian far from criticality;
and the mean-field critical picture would predict a PDF proportional to $\exp(-c|A|^{4})$. 
(iii) {\it What can we say about the power spectral density (PSD) of $|A(t)|$?}
In particular, how does it scale with $N$?
(iv) {\it What is the width of the critical region as a function of $N$?}

\begin{figure}[htbp]
  \centering
  \includegraphics[width=0.8\linewidth]{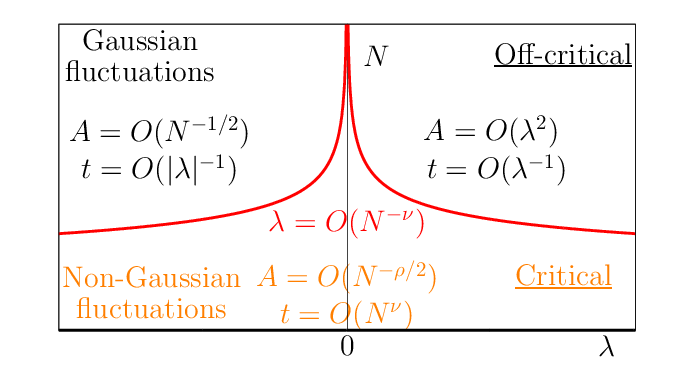}
  \caption{
    Schematic figure of scalings around marginal stability.
    $A$ is an order parameter, $t$ is the time, and $\lambda$ is the real growth rate.
    Critical scaling of finite size fluctuations is crucial in the critical region.
    $\rho\simeq 4/5$ and $\nu\simeq 1/5$. (Non-)Gaussian fluctuations are for $A(t)$.
  }
  \label{fig:Schematic}
\end{figure}

In this article, we introduce a phenomenological description of the fluctuations,
which answers the above four questions.
The first result is the scaling $A\propto N^{-2/5}$,
which implies the exponent $\rho=4/5$ as observed in simulations \cite{Yoshi2016}.
As a second result, we predict a universal form for the probability distribution
of the order parameter at the critical point,
which is neither Gaussian, nor related to the critical mean-field picture.
As a third result, our description naturally induces
the time scaling $t\propto N^{\nu}\tau$ with $\nu=1/5$.
Then the PSD of the order parameter, denoted by $S(f)$,
admits a universal scaling form:
$N^{\rho-\nu} S(N^{\nu}f)$ must fall on a universal curve independent of $N$.
As a fourth result we introduce a scaled growth ratio
$\epsilon=N^{\nu}\lambda$ which is induced by the time scaling.
This parameter determines the boundary between the critical regime
dominated by finite $N$ fluctuations $|A| \sim N^{-\rho/2}$
and the far from critical regime dominated by the standard trapping scaling picture
$|A|\sim \lambda^{2}$, decorated by small finite size fluctuations.
Comparisons with extensive direct numerical simulations of
two simple models confirm the validity of these predictions,
including the approximate values for the exponents $\rho=4/5$ and $\nu=1/5$.
See Fig.~\ref{fig:Schematic} for a summary.
We expect that these results describe finite $N$ effects
close to any instance of the SWM bifurcation.

To complete our comparison with the existing literature,
we note that finite $N$ fluctuations are studied in a version of vector resonant relaxation,
a collisionless model, in \cite{Fouvry2019,Flores25,Flores25b},
but there is no bifurcation in this case and the scaling is not investigated.
Critical fluctuations have also been studied
in the context of Kuramoto model of synchronizing oscillators.
However, in \cite{ChatePRL,ChatePRE},
the non mean-field exponent describing the scaling with $N$
of the variance of the order parameter refers to fluctuations in the quenched disorder,
whereas we are interested in the dynamical fluctuations for a single realization.

\paragraph{A phenomenological equation for the order parameter.}
The SWM bifurcation requires the following conditions:
i) a single mode becomes unstable at the bifurcation point,
ii) the bifurcating mode is in resonance with the continuous spectrum of the linearized dynamics,
and iii) the cat's eye resulting from this resonance has a width of order $|A|^{1/2}$,
where $A$ is the complex mode's amplitude.
We remark that a single mode bifurcation with resonance can be in another universality class \cite{CrawfordJarayanan}.

We thus consider $N$ interacting particles,
such that the associated Vlasov equation has only one unstable mode near the stability threshold.
The amplitude of this mode $A(t)$ will be our main object of study.
For dissipative systems, center manifold theory allows to write
an autonomous differential equation for $A(t)$ \cite{HaragusIoos};
when finite $N$ fluctuations are taken into account,
one expects this equation to be supplemented by a noise term
(see \cite{Comets88,Collet17} for rigorous proofs in special cases). 
However, such an autonomous equation for $A$ does not exist for the SWM  bifurcation:
an infinite dimensional weakly nonlinear dynamics remains \cite{DelCastilloNegrete,BMT}.

In order to gain insight into the critical regime,
{\it a self-consistent equation},
relating the initial state and the final state close to the bifurcation,
has been introduced in \cite{Leoncini09,deBuyl11,YoshiShun14,YoshiShun15}.
Although it is not a dynamical equation
it is very successful in describing asymptotic behavior
around the critical point \cite{YoshiShun15b};
see also \cite{Teles25}. It reads
\[
  -\Lambda(0) A+L_{3/2}A|A|^{1/2} + O(|A|^{2}) =0, 
\]
where $\Lambda(\lambda)$ is the spectral function:
$\Lambda(\lambda)=0$ for $\lambda$ eigenvalue of the linearized operator.
This equation relies on 
the cat's eye structure in phase space and the conservation of Casimirs. 

In particular, the dominant nonlinear term $O(|A|^{3/2})$ arises
because the cat's eye width in the momentum direction is of order $|A|^{1/2}$,
and we expect that a dynamical equation for the order parameter $A(t)$
would have the same type of leading nonlinearity. Based on these premises,
we introduce a phenomenological equation:
\begin{eqnarray}
  \dfrac{dA}{dt}
  &=&\lambda A -c_{3/2}A|A|^{1/2} +
      \frac{\sigma}{\sqrt{N}}\eta(t),
      \label{eq:pheno}
\end{eqnarray}
where $\eta$ 
is an independent Gaussian delta correlated white noise
crudely modeling the Poisson shot noise due to finite but large $N$,
$\lambda$ is the bifurcating eigenvalue,
$c_{3/2}$ is the coefficient of the dominant nonlinear term
(which could be fitted knowing the saturation amplitude of $A$
at positive $\lambda$ and infinite $N$),
and $\sigma$ controls the finite size noise intensity.

Equation \eqref{eq:pheno} is the foundation of our results.
We do not expect this equation to be quantitatively accurate,
rather we aim at describing qualitatively the behavior of $A(t)$,
namely scaling laws, close to the bifurcation point.
Rescaling the amplitude $A$ and the time $t$ as
$A=[\sigma^{2}/(c_{3/2}N)]^{2/5}a$ and
$t=N^{1/5}/(c_{3/2}^{2}\sigma)^{2/5}\tau$,
Eq. \eqref{eq:pheno} becomes
\begin{equation}
  \frac{da}{d\tau}
  = \widetilde{\epsilon} a -  a |a|^{1/2} + \widetilde{\eta}(\tau),~
  ~ \widetilde{\epsilon} = \dfrac{\epsilon}{(c_{3/2}^{2}\sigma)^{2/5}},~
  \epsilon = N^{1/5}\lambda,
  \label{eq:scaled_pheno}
\end{equation}
where the noise is also scaled as $\eta(t)\sqrt{dt}=\widetilde{\eta}(\tau)\sqrt{d\tau}$.
From Eq.~\eqref{eq:scaled_pheno}, we obtain the four results mentioned above:
the variance $V\propto N^{-\rho}$;
the anomalous PDF with $\exp(-|a|^{\mu})$ tails for $\epsilon=0$;
the universal PSD $N^{\rho-\nu}S(N^{\nu}f)$;
and the scaled growth ratio $\epsilon=N^{\nu}\lambda$,
where $\rho=4/5$, $\mu=5/2$, and $\nu=1/5$.
In Appendix A, we show that the unstable manifold expansion
developed by Crawford \cite{Crawford94,Crawford95}
yields the same scalings for amplitude, time, and critical region,
although it cannot access the critical point.
We emphasize that these predictions rely only on the hypothesis
of a SWM bifurcation at the continuous level.
Hence they are in principle very generic. 
We now turn to two specific examples to test them.

\paragraph{Hamiltonian mean-field model.}
As the simplest testbed for our predictions,
we use the Hamiltonian mean-field model whose
$N-$body Hamiltonian is
\begin{equation}
  H_{\rm HMF} = \sum_{i=1}^{N}\frac{p_{i}^{2}}{2} + \frac{1}{2N} \sum_{i=1}^{N} \sum_{j=1}^{N} \big[1-\cos(q_{i}-q_{j})\big].
  \label{eq:HMF}
\end{equation}
It is a paradigmatic toy model for long range interacting systems \cite{ReviewLongrange},
but essentially the same model has also been introduced in an astrophysical context
\cite{Inagaki93,PichonHMF}.
Initially, positions $\{q_{i}\}$ are taken uniformly in $[0,2\pi)$, and momenta
$\{p_{i}\}$ distributed according to the distribution
$F_{0}(p)=\sqrt{\beta/(2\pi)} e^{-\beta p^2/2}$,
normalized as $\int_{\mathbb{R}} F_{0}(p) dp=1$.
In the $N\to \infty$ limit, the dynamics of the particles distribution
is governed by a Vlasov equation, and $\beta_{\rm c}=2$ corresponds to the bifurcation point.
The bifurcating eigenvalue $\lambda$,
the growing rate, is proportional to $\beta_{\rm c}-\beta$ around this point.
The magnetization 
$\mathbf{M}=(M_{x},M_{y})=\sum_{j}(\cos q_{j},\sin q_{j})/N$
plays the role of the complex order parameter $A$;
we write $M=|\mathbf{M}|$.
All simulations are performed by using a fourth-order symplectic integrator \cite{Yoshida90}
with the time step $\Delta t=0.1$.
 
\begin{figure}[htbp]
  \centering
  \includegraphics[width=1.0\linewidth]{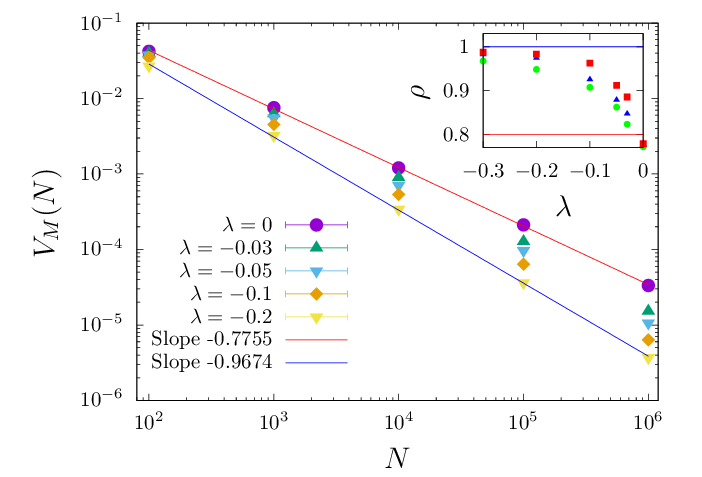}
  \caption{
    Variance $V_{M}$ computed for $t\in [20,100]$
    on the stable side of the bifurcation, for the HMF model:
    $\lambda=0, -0.03, -0.05, -0.1,$ and $-0.2$ from top to bottom.
    Error bars, the standard deviation over $100$ realizations,
    are smaller than the symbols.
    The guide lines are computed by the least square method for $\lambda=0$ (red, slope $-0.7755$)
    and $\lambda=-0.2$ (blue, slope $-0.9674$) for $N\in [10^{2},10^{6}]$.
    Inset shows $\lambda$ dependence of $\rho$ corresponding to slopes,
    which are computed in the interval $N\in [10^{2},10^{4}]$ (green circles),
    $[10^{3},10^{5}]$ (blue triangles), and $[10^{4},10^{6}]$ (red squares).
    The red and blue horizontal lines mark $\rho=4/5$ and $1$ respectively.
  }
  \label{fig:HMF_Variance}
\end{figure}

As a first test of Eq.~\eqref{eq:scaled_pheno}, we study the variance of $M$ in the stable region.
We compute it as $V_{M}=\ave{M^{2}}_{t}$,
where $\ave{\cdot}_{t}$ represents the time average, for $t\in [20,100]$
to avoid an initial transient regime.
We then average over $100$ realizations
and report the result in Fig.~\ref{fig:HMF_Variance}.
The best fit is $V_{M}(N)\sim N^{-0.7755}$ at the critical point,
where the exponent is close to the theoretically predicted value $\rho=4/5$.
The exponent approaches the normal value $\rho=1$
as the system goes away from the critical point.
We remark that the exponent is closer to $4/5$ if estimated in a smaller $N$ interval
(see the inset of Fig.~\ref{fig:HMF_Variance}).
This tendency is consistent with a critical region's width governed
by the universal parameter $\epsilon=N^{1/5}\lambda$
(see Fig.~\ref{fig:Schematic}),
which will be precisely examined later.
In Appendix B, we also compute the variance of $\dot{M}_{x}$,
and check that the CLT-type scaling of the noise in Eq.~\eqref{eq:pheno} is good.

\begin{figure}[htbp]
  \centering
  \includegraphics[width=1.0\linewidth]{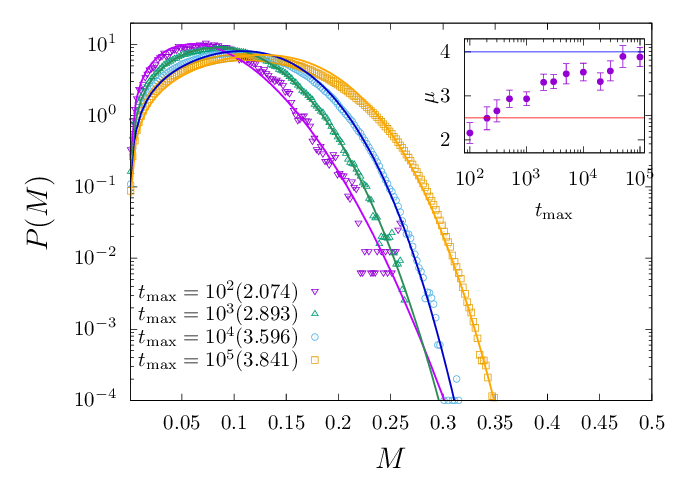}
  \caption{Probability distribution function $P(M)$ for $N=10^{3}$ at $\lambda=0$
    in the HMF model. PDFs are averaged for $t\in [20,t_{\rm max}]$
    and $10^{2}$ realizations.
    $t_{\rm max}=10^{2},10^{3},10^{4},10^{5}$ from top left to bottom right;
    in parentheses is the estimated value of $\mu$.
    Inset shows $t_{\rm max}$ dependence of the exponent $\mu$,
    which is the average (points) and standard deviation (error bars) over $10^{3}$ resamples
      by the bootstrap method (see Appendix D).
    The red and blue horizontal lines mark $\mu=5/2$ and $4$ respectively.
  }
  \label{fig:HMF_Dist}
\end{figure}

The second test is the distribution.
On Fig.\ref{fig:HMF_Dist} is plotted the PDF of the magnetization modulus $M$,
at the critical point, averaged from $t=20$ to $t=t_{\rm max}$ with $N=10^{3}$.
The PDF is approximated by $P(M)=cM\exp(-bM^{\mu})$,
where the factor $M$ comes from the Jacobian for the two-dimensional magnetization
$\mathbf{M}$,
and $\mu,b$, and $c$ are obtained by the least square mean method for $\log P(M)$
(see Appendix C).
We focus on the exponent $\mu$.
Clearly, for small $t_{\rm max}$ the exponent is close to $\mu=5/2$,
while for large $t_{\rm max}$, the expected exponent $\mu=4$ is observed,
since the system has at least partially relaxed to full statistical equilibrium.

\begin{figure}[ht]
  \centering
  \includegraphics[width=1.0\linewidth]{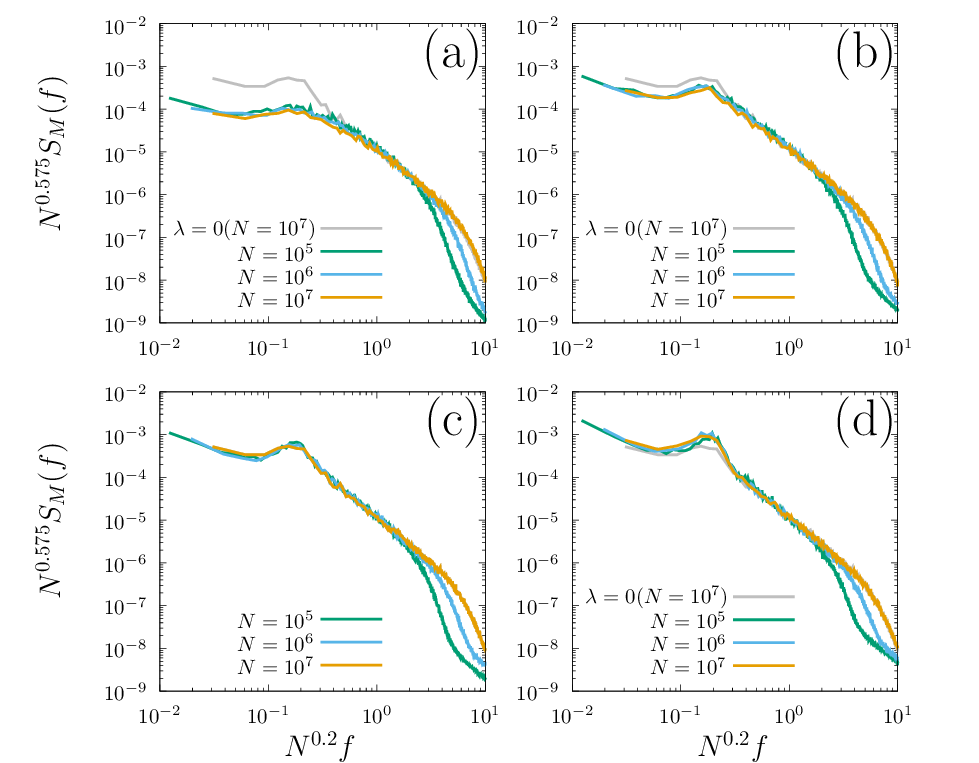}
  \caption{
    Scaled power spectral density $N^{\rho-\nu}S_{M}$ as a function of $N^{\nu}f$ in the HMF model.
    $\rho=0.775, \nu=0.2$.
    The parameter $\epsilon=N^{\nu}\lambda$ is
    (a) $\epsilon=-1$, (b) $-0.3$, (c) $0$, and (d) $0.3$.
    $N=10^{5}$ (green), $10^{6}$ (blue), and $10^{7}$ (orange).
    The gray curve is $N^{\nu}\lambda=0$ with $N=10^{7}$ for comparison.
    Average over $100$ realizations.
    Time series are taken in the time region $t\in [180.9, 1000]$
    to avoid an initial transient regime and for convenience of the fast Fourier transform.
  }
  \label{fig:HMF_PSD}
\end{figure}

The third and fourth results are tested on the PSD $S_{M}(f)$ of $M(t)$, 
varying the scaled growth ratio $\epsilon=N^{1/5}\lambda$.
See Appendix E for the PSD.
The PSD averaged over $100$ realizations is shown in Fig.~\ref{fig:HMF_PSD};
for small frequency $f$ and at fixed $\epsilon$, PSDs fall on an $N$-independent curve; 
and the frequency scaling $N^{1/5}f$ agrees with
the predicted time scaling $t\propto N^{1/5}\tau$.

\paragraph{Two-dimensional Euler-like model.}
Two-dimensional Euler equation shares similar mathematical properties with Vlasov equation,
and indeed it does present SWM bifurcations.
However, the underlying particles dynamics in this case involves point vortices:
the corresponding Hamiltonian has no kinetic energy,
and thus is quite different from the standard Vlasov case.
In order to test the universality of our predictions,
we now turn to a simplified version of interacting point vortices.   

The model inspired by the two-dimensional Euler fluid is
\begin{equation}
  H_{\rm Euler} = -\dfrac{1}{N} \sum_{i=1}^{N} \sum_{j=1}^{N}
  [ \cos(2\pi (q_{i}-q_{j})) + \cos(2\pi(p_{i}-p_{j})) ].
  \label{eq:Euler}
\end{equation}
Initially, $\{q_{i}\}$ are taken uniformly in $[0,1)$,
and $\{p_{i}\}$ distributed according to
$G_{0}(p)=1+(1/2)\cos(2\pi p)+g\cos(6\pi p)$
normalized as $\int_{0}^{1} G_{0}(p) dp = 1$.
The parameter $g$ is the bifurcation parameter:
the critical point is $g_{\rm cr}=0$ and $g<0$ ($g>0$) implies stability (instability):
this is a caricature of shear flow instability.
The vector $\mathbf{W}=\sum_{j}(\cos(2\pi q_{j}),\sin(2\pi q_{j}))/N$ plays the role of $A$;
we write $W=|\mathbf{W}|$.
See Appendix F for the computation of the spectral function $\Lambda(\lambda)$ in this case.

\begin{figure}[htbp]
  \centering
  \includegraphics[width=1.0\linewidth]{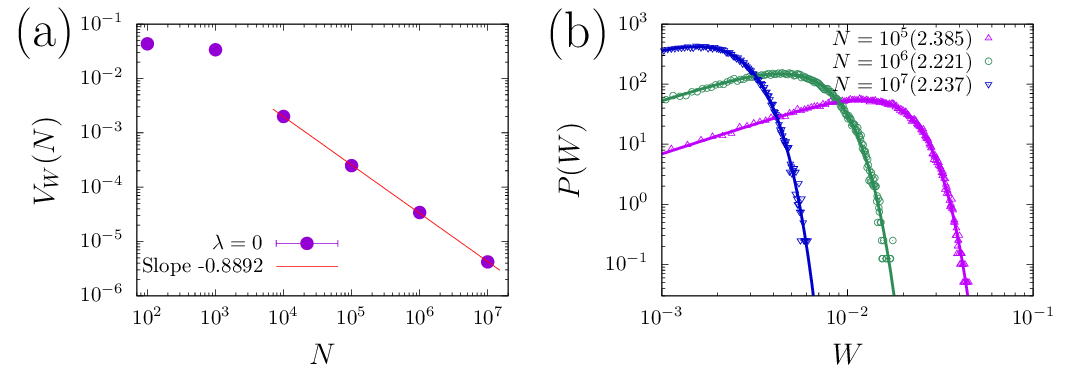}
  \caption{Test of the phenomenological equation \eqref{eq:scaled_pheno}
    at the critical point in the Euler-like model Eq.~\eqref{eq:Euler},
    computed from $t\in [20,100]$ of $100$ realizations.
    (a) Variance of $W$ at the critical point.
    The red line represents the best fit for the four points in $N\in [10^{4},10^{7}]$.
    The slope is $-0.8892$ while the prediction is $-\rho=-4/5$.
    (b) Probability distribution function for $N=10^{5}, 10^{6}$, and $10^{7}$
    from bottom right to top left; in parentheses is the estimated value of $\mu$ whose theoretical value is $\mu=5/2$.
  }
  \label{fig:Euler_Variance_Dist}
\end{figure}

\begin{figure}[htbp]
  \centering
  \includegraphics[width=1.0\linewidth]{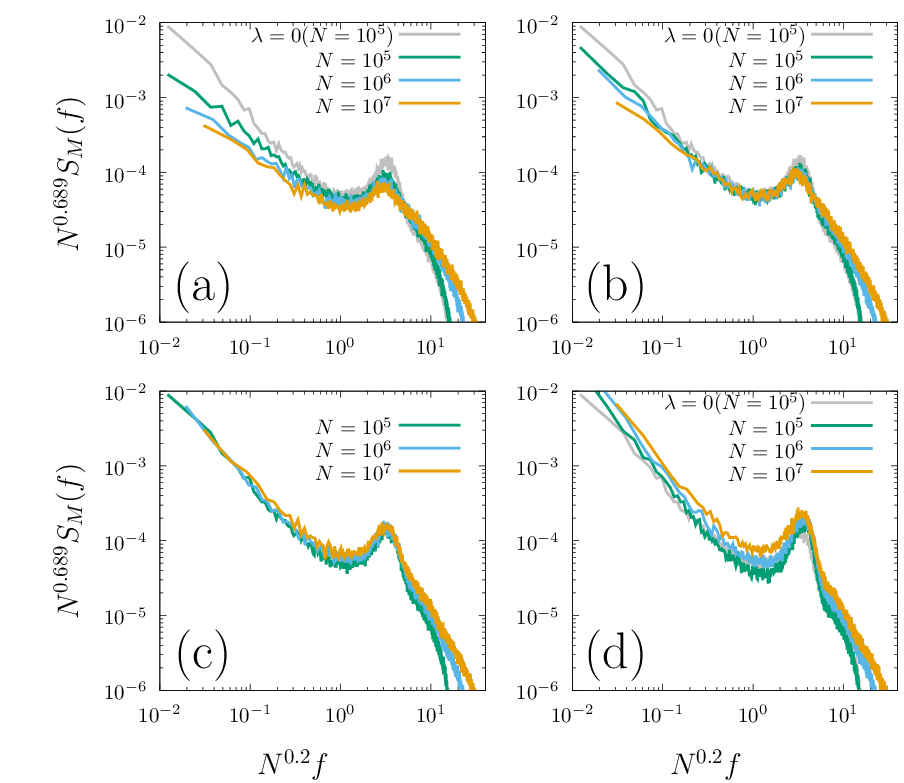}
  \caption{Same as Fig.~\ref{fig:HMF_PSD} but in the Euler-like model Eq.~\eqref{eq:Euler}.
    The parameter $\epsilon=N^{\nu}\lambda$ is
    (a) $\epsilon=-10$, (b) $-5$, (c) $0$, and (d) $5$.
  }
  \label{fig:Euler_PSD}
\end{figure}

At the critical point the variance $V_{W}=\ave{W^{2}}_{t}$
and PDFs $P(W)$ are shown in Fig.~\ref{fig:Euler_Variance_Dist}.
The fitting curves for $P(W)$ assume $P(W)=cW\exp(-bW^{\mu})$
as in the HMF model.
They are in good agreement with the theoretical predictions.
Universality of PSDs of $W(t)$ is examined in Fig.~\ref{fig:Euler_PSD}.
As observed in the HMF model (see Fig.~\ref{fig:HMF_PSD}),
universality holds around the critical point.

\paragraph{Conclusion}
Just like SWM bifurcation itself,
we expect our findings to have a wide validity across different physical fields.
This work suggests many further numerical studies of critical fluctuations
in various more realistic contexts, such as self-gravitating systems close to Jeans instability,
or interacting point vortices.
Our results also open the door to many new theoretical investigations.
First, while the numerical computations of the critical exponents
from direct $N$-body simulations are in good agreement with the phenomenologically predicted values,
the theory is only approximate,
and the question of the exact value of the critical exponents remains open.
The phenomenological equation \eqref{eq:pheno}
also does not explain the bump and $1/f$-like spectrum (see also \cite{Yoshi18}) observed in $N$-body simulations of the PSD.
Attacking these problems could rely on the derivation of a "discrete Single Wave Model",
possibly following the ideas of \cite{ElskensBook,Escande18};
or using field theory \`a la Martin-Siggia-Rose,
as done recently for Vector Resonant Relaxation in \cite{Flores25}. 
Furthermore, despite the importance of the SWM bifurcation,
it is not the only type of bifurcation occurring in Vlasov systems,
see \cite{CrawfordJarayanan,BMY20,YB23,YB25} for other examples.
What do the finite $N$ fluctuations look like close to these bifurcations?  

The implications of our findings for kinetic theory is also to be investigated, in particular in the context of self-gravitating systems, where collisional relaxation close to marginal stability is  an open question \cite{Fouvry}.

\acknowledgements
Y.Y.Y. acknowledges the support of JSPS KAKENHI Grant No. 21K03402,
and J.B. the support of  ANR project RETENU ANR-20-CE40-0005-01.

\clearpage
\appendix

\begin{center}
    \textbf{End Matter}
\end{center}
\vspace{-0.3cm}
\setcounter{equation}{0}

\section{Appendix A : Unstable manifold expansion}
\label{sec:Crawford}
A finite dimensional unstable manifold expansion for the SWM bifurcation,
which is valid in the weakly unstable regime,
has been obtained by Crawford \cite{Crawford94,Crawford95}.
However, it comes in the form of an infinite series,
with coefficients diverging at the critical point.
It is natural to add heuristically a noise term to this expansion,
in order to mimic the effect of finite size fluctuations.
The reduced Crawford equation then reads
\begin{equation}
  \dfrac{dA}{dt} = \lambda A - \sum_{k\geq 1}c_{2k+1}(\lambda) A|A|^{2k} +\frac{\sigma}{\sqrt{N}}\eta(t),
\label{eq:crawford_not_rescaled}
\end{equation}
where $A\in\mathbb{C}$,
$\eta(t)$ is a standard delta-correlated white noise.
This expansion is singular at small $\lambda$,
as $c_{2k+1}(\lambda) \sim C_{2k+1}/\lambda^{4k-1}$.
Rescaling \eqref{eq:crawford_not_rescaled} using $A=\lambda^{2}a, t=\lambda^{-1}\tau$,
and $\eta(t)\sqrt{dt}=\tilde{\eta}(\tau)\sqrt{d\tau}$,
truncating at order $a^{3}$,
and keeping the leading order,
we obtain the rescaled equation:
\begin{equation}
  \dfrac{da}{d\tau}
  = a-C_{3}|a|^{2} a +(N\lambda^{5})^{-1/2}\sigma \tilde{\eta}(\tau).
\end{equation}
One can then distinguish two regimes.
$N\lambda^{5}\gg 1$, far from criticality:
the finite size fluctuations act as a small perturbation on the usual bifurcation structure;
$N\lambda^5 \ll 1$, critical regime:
the finite size fluctuations dominates the dynamics.
Equation \eqref{eq:crawford_not_rescaled} does not allow to investigate the critical regime.
However, approaching the crossover between the two regimes,
$N\lambda^{5} \simeq 1$ corresponds to $|A|\simeq N^{-\rho/2}$ with $\rho=4/5$,
since trapping scaling is $|A|=O(\lambda^{2})$.
Hence Crawford's expansion, complemented by a stochastic term, recovers some of our results.

\section{Appendix B : Test of the noise size in Eq.~\eqref{eq:pheno}}
\label{sec:noise-size}

We focus on the critical point $\lambda=0$.
The phenomenological equation is:
\begin{equation}
  \dfrac{dA}{dt} = f(A) + \dfrac{\sigma}{N^{\zeta}} \eta(t),
  \qquad
  f(A) = -c_{3/2} A|A|^{1/2},
\end{equation}
where we introduced the exponent $\zeta$ to examine if $\zeta=1/2$.
The variance of $dA/dt$ denoted by $V_{\dot{A}}$ is
\begin{equation}
  V_{\dot{A}} = V_{f} + \dfrac{2\sigma}{N^{\zeta}} \ave{f(A)\eta(t)} + \dfrac{\sigma^{2}}{N^{2\zeta}} \ave{\eta(t)^{2}},
\end{equation}
where $\ave{\cdots}$ represents the ensemble average and $V_{f}=\ave{f(A)^{2}}-\ave{f(A)}^{2}$.
Remembering $A=O(N^{-2/5})$ at the critical point and
picking up the largest term, we conclude
\begin{equation}
  V_{\dot{A}} = O(1/N) \quad\Longrightarrow\quad \zeta=1/2. 
\end{equation}

\begin{figure}[htbp]
  \centering
  \includegraphics[width=1.0\linewidth]{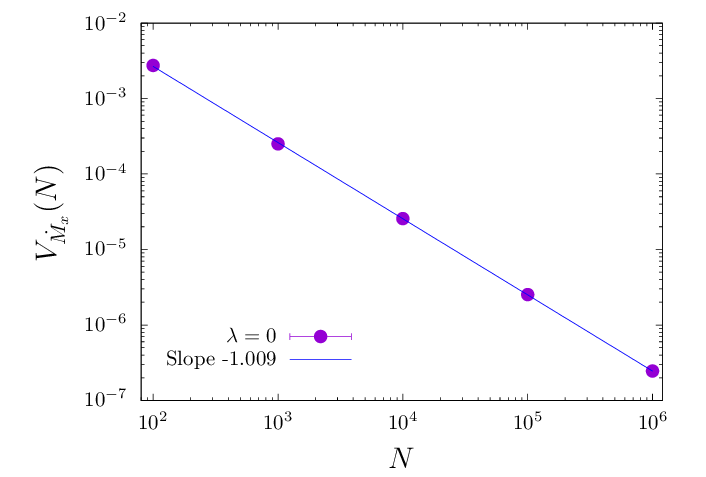}
  \caption{
    Variance of $dM_{x}/dt$ at the critical point in the HMF model.
    The slope $-1.009$ is computed in the interval $N\in [10^{2},10^{6}]$
    by using the least square method.}
  \label{fig:HMF_VarianceMdot}
\end{figure}

In the HMF model,  $M_{x}$ plays the role of the real part of $A$ and its explicit form is
\begin{equation}
  \dfrac{dM_{x}}{dt} = - \dfrac{1}{N} \sum_{j=1}^{N} p_{j} \sin(q_{j}).
\end{equation}
Therefore, $V_{\dot{M}_{x}}=O(1/N)$ implies $\zeta=1/2$.
This is indeed true as shown in Fig.~\ref{fig:HMF_VarianceMdot}.

\section{Appendix C: Fitting of probability distribution function}
\label{sec:Fitting}

Let $\{(M_{i},P_{i})\}_{i=1}^{n}$ represent a numerically obtained discrete PDF.
We fit these points by the logarithm of $P(M)=cM\exp(-bM^{\mu})$
by searching the parameter set $(\mu,b,c)$ which minimizes the error function
\begin{equation}
  L(\mu,b,c) = \dfrac{1}{2} \sum_{i=1}^{n}
  \left[ \ln P_{i} - \left( \ln c + \ln M_{i} - b M_{i}^{\mu} \right) \right]^{2}.
\end{equation}
Let $\ave{\cdots}$ denote the $n$ point average,
i.e. $\ave{M_{i}}=\sum_{i=1}^{n}M_{i}/n$ for instance.
The minimizing condition, $\nabla L=0$, gives
\begin{equation}
  \ln c = \ave{\ln P_{i}} - \ave{\ln M_{i}} + b \ave{M_{i}^{\mu}},
  \label{eq:lnc}
\end{equation}
\begin{equation}
  b = - \dfrac{\ave{M_{i}^{\mu}(\ln P_{i}-\ln M_{i})}
    - \ave{M_{i}^{\mu}}\left( \ave{\ln P_{i}} - \ave{\ln M_{i}} \right) }
  { \ave{M_{i}^{2\mu}} - \ave{M_{i}^{\mu}}^{2}},
  \label{eq:b}
\end{equation}
and
\begin{equation}
  \begin{split}
    & \ave{M_{i}^{\mu}\ln M_{i}\ln P_{i}} - \ln c \ave{M_{i}^{\mu}\ln M_{i}} \\
    & - \ave{M_{i}^{\mu}(\ln M_{i})^{2}} + b \ave{M_{i}^{2\mu}\ln M_{i}} = 0.
  \label{eq:mu}
  \end{split}
\end{equation}
Substituting Eqs.~\eqref{eq:lnc} and \eqref{eq:b} into Eq.~\eqref{eq:mu},
we have a nonlinear equation for the variable $\mu$,
whose solution gives the minimizing parameter set.

\section{Appendix D: Bootstrap method}

We compute $10^{2}$ time series (realizations) of $M(t)$
and pick up the time interval $[20,t_{\rm max}]$ from each time series.
First, we uniformly randomly choose $10^{2}$ time series
from the computed $10^{2}$ time series allowing duplication.
Second, by following Appendix C, we compute the best fit parameter set of PDF $P(M)$
constructed from the resampled $10^{2}$ time series.
Finally, we repeat the resampling procedure $10^{3}$ times,
and compute the average and standard deviation of the exponent $\mu$.

\section{Appendix E: Power spectral density}
\label{sec:PSD}

Let $F(t)$ be a time series and $F_{n}=F(n\Delta t)~(n=0,\cdots,N-1)$ be a discrete sampling.
We compute the Fourier components $\widetilde{F}_{k}~(k=0,\cdots,N-1)$ by
\begin{equation}
  \widetilde{F}_{k} = \dfrac{1}{N} \sum_{n=0}^{N_1} e^{-i \frac{2\pi}{N}kn} F_{n}.
\end{equation}
This transformation can be performed by using the fast Fourier transform
if $N=2^{n}$ with $n\in\mathbb{N}$.
The corresponding frequency is $f_{k}=k/(N\Delta t)$,
and the power spectral density $S(f)$ is defined by
\begin{equation}
  S(f_{k}) = | \widetilde{F}_{k}|^{2}.
\end{equation}

\section{Appendix F: Spectrum function in the Euler-like model}
\label{sec:Lambda-EulerHMF}

The linearized equation gives the spectrum function for the Fourier mode $1$ in the direction $q$, for ${\rm Re}(\lambda)>0$:
\begin{equation}
  \begin{split}
    \Lambda(\lambda)
    & = 1 - \dfrac{1}{2\pi} \int_{-\pi}^{\pi} \dfrac{\sin(p)+6g\sin(3p)}{\sin(p)-i\lambda/\pi^{2}} dp.
  \end{split}
\end{equation}
We have $\Lambda(0)=0$ at $g=0$, and $g=0$ is the critical point.
Apart from $\lambda=0$, the integrand is singular for ${\rm Re}(\lambda)=0$,
and we analytically continue $\Lambda(\lambda)$ from ${\rm Re}(\lambda)>0$
to ${\rm Re}(\lambda)\leq 0$ by following the Landau's recipe:
picking up residues coming from poles crossing the integral contour at ${\rm Re}(\lambda)=0$
and adding them to $\Lambda(\lambda)$.
We have two poles satisfying $\sin(p_{\ast})=i\lambda/\pi^{2}$,
and around the critical point,
one located around $p_{\ast}\simeq 0$ approaches the real axis from the upper-half $p$-plane
and the other located around $p_{\ast}\simeq \pi$ from the lower-half.
The pole coming from the upper-half provides the residue
\begin{equation}
  i 2\pi \dfrac{-1}{2\pi} \dfrac{\sin(p_{\ast})+6g\sin(3p_{\ast})}{\cos(p_{\ast})}
\end{equation}
for ${\rm Re}(\lambda)<0$ and the other have the opposite sign.
Using $\sin(p_{\ast})=i\lambda/\pi^{2}$,
the total residue contribution to $\Lambda(\lambda)$ is
\begin{equation}
  R(\lambda) = \dfrac{2\lambda/\pi^{2}}{\sqrt{1+(\lambda/\pi^{2})^{2}}}
  \left\{ 1 + 6 g \left[ 3 + 4 (\lambda/\pi^{2})^{2} \right] \right\}
\end{equation}
for ${\rm Re}(\lambda)<0$ and $R(\lambda)/2$ for ${\rm Re}(\lambda)=0$.

\clearpage

\onecolumngrid

\setcounter{equation}{0}
\setcounter{section}{0}

\end{document}